\documentclass[conference]{IEEEtran}
\IEEEoverridecommandlockouts

\usepackage{hyperref}
\hypersetup{hidelinks}
\usepackage{cite}
\usepackage{amsmath,amssymb,amsfonts,mathtools}
\usepackage{algorithmic}
\usepackage{graphicx}
\usepackage{textcomp}
\usepackage{xcolor}

\usepackage{amsthm}

\def\BibTeX{{\rm B\kern-.05em{\sc i\kern-.025em b}\kern-.08em
    T\kern-.1667em\lower.7ex\hbox{E}\kern-.125emX}}
\begin{document}

\title{Dynamic Complex-Frequency Control of Grid-Forming Converters\\
\thanks{This work was supported by the European Union's Horizon 2020 and 2023 research and innovation programs (Grant Agreement Numbers 883985 and 101096197).}
}

\author{
\IEEEauthorblockN{Roger Domingo-Enrich\IEEEauthorrefmark{1}\IEEEauthorrefmark{2}, Xiuqiang He\IEEEauthorrefmark{2}, Verena Häberle\IEEEauthorrefmark{2}, Florian Dörfler\IEEEauthorrefmark{2}}%
\IEEEauthorblockA{\IEEEauthorrefmark{1} Universitat Politècnica de Catalunya, 08028 Barcelona, Spain } %
\IEEEauthorblockA{\IEEEauthorrefmark{2} Automatic Control Laboratory, ETH Zurich, 8092 Zurich, Switzerland \\
Emails: \{rdomingo,xiuqhe,verenhae,dorfler\}@ethz.ch}
}

\maketitle

\begin{abstract}
Complex droop control, alternatively known as dispatchable virtual oscillator control (dVOC),  stands out for its unique capabilities in synchronization and voltage stabilization among existing control strategies for grid-forming converters. Complex droop control leverages the novel concept of ``complex frequency'', thereby establishing a coupled connection between active and reactive power inputs and frequency and rate-of-change-of voltage outputs. However, its reliance on static droop gains limits its ability to exhibit crucial dynamic response behaviors required in future power systems. To address this limitation, this paper introduces \textit{dynamic complex-frequency control}, upgrading static droop gains with dynamic transfer functions to enhance the richness and flexibility in dynamic responses for frequency and voltage control. Unlike existing approaches, the complex-frequency control framework treats frequency and voltage dynamics collectively, ensuring small-signal stability for frequency synchronization and voltage stabilization simultaneously. The control framework is validated through detailed numerical case studies on the IEEE nine-bus system, also showcasing its applicability in multi-converter setups.
\end{abstract}


\section{Introduction}
The widespread integration of distributed generation into power grids is displacing centralized synchronous generators, traditionally responsible for grid-forming (GFM) tasks such as frequency and voltage regulation. This transition is prompting the delegation of GFM responsibilities to distributed generation systems. Consequently, GFM converters are emerging as vital components in modern power systems, especially in facilitating the provision of dynamic ancillary services \cite{musca2022grid}. Various GFM converter control strategies have been proposed and extensively investigated, including classical droop control \cite{simpson2013synchronization,ainsworth2013structure}, virtual synchronous machine (VSM) \cite{shuai2018transient}, dispatchable virtual oscillator control (dVOC) \cite{colombino2019global,seo2019dispatchable,lu2022virtual}, and matching control \cite{arghir2019electronic}. Among these strategies, dVOC has gained increasing attention due to its unique stability performance in synchronization and voltage stabilization.

Our recent work \cite{he2023nonlinear} has reformulated dVOC equivalently as complex-power complex-frequency (\textit{complex droop control} for brevity), leveraging a novel concept known as ``complex frequency'' \cite{milano2021complex}. Complex frequency offers a unified and concise representation of angular frequency and rate-of-change-of-voltage. Accordingly, complex droop control links active and reactive power inputs to frequency and rate-of-change-of-voltage outputs, introducing non-trivial coupling. As a result, complex droop control is inherently multivariable and nonlinear, contrasting with single-input single-output linear loops in classical droop control \cite{simpson2013synchronization,ainsworth2013structure} or VSM \cite{shuai2018transient}. While complex droop control effectively manages the inherent coupling and nonlinearity in active and reactive power flows, its capability to provide grid ancillary services for frequency and voltage regulation is limited to static droop gains exclusively. It thus lacks the adaptability to deliver a \textit{rich and dynamic} response behavior, such as offering inertial response or fast dynamic ancillary services, which are crucial for future power systems.

In this paper, we propose the concept of dynamic complex-frequency control to incorporate rich dynamism. Specifically, we advocate upgrading static droop gains with dynamic gains, represented by dynamic transfer functions. This upgrading offers sufficient richness and flexibility in providing arbitrarily desired frequency and voltage dynamic responses. Moreover, our proposed framework, termed ``dynamic complex-frequency control'', holds promise for application in multi-converter scenarios, where multiple units can collectively achieve desired dynamic complex-frequency control behaviors. This approach accounts for the individual unit limitations in timescale and capacity, akin to the recent concept for dynamic virtual power plants \cite{bjork2022dynamic,haberle2021control,haberle2023grid}. Furthermore, in contrast to the existing work in \cite{jiang2021grid,haberle2023grid}, which examines multi-converter setups for grid-forming dynamic ancillary services provision based on classical decoupled frequency and voltage controls, our framework, rooted in complex-frequency coordinates and collectively treating frequency and voltage controls, effectively manages the coupling between active power-frequency and reactive power-voltage dynamics. Consequently, it enables us to offer more comprehensive small-signal stability guarantees for both frequency synchronization and voltage stabilization simultaneously. Specifically, in contrast to the results in \cite{jiang2021grid,paganini2019global}, where the closed-loop stability is established under the assumption of fixed voltages, our stability analysis inherently accounts for variable frequency and voltage dynamics.

The remainder of this paper is organized as follows. Section \ref{sec:preliminaries} outlines preliminaries concerning the concepts of complex frequency, complex power, and static complex droop control. In Section \ref{sec:dynamic_complex_freq_control}, we introduce a novel framework for dynamic complex-frequency control. The efficacy of dynamic complex-frequency control is illustrated through numerical case studies in Section \ref{sec:case_studies}. Finally, Section \ref{sec:conclusion} concludes the paper.

\section{Preliminaries}\label{sec:preliminaries}
In this section, we first recall the definition of complex frequency \cite{milano2021complex} and complex power \cite{he2024complex} and then introduce the static complex droop control\cite{he2023nonlinear}.

\subsection{Complex Frequency}
We consider a single-phase or balanced three-phase voltage-source converter, which is modeled as a voltage source since we apply GFM control. We can express the terminal voltage of the converter in terms of a complex variable as
\begin{equation}\label{compl_volt}
\underline{v}=v(\cos{\theta}+j\sin{\theta})=ve^{j\theta},
\end{equation}
where ${v}$ is the associated voltage amplitude and $\theta$ the phase angle. Using the voltage amplitude logarithm $u=\ln{v}$ ($v$ in per unit) and the phase angle $\theta$ (in radians), we define the complex angle and its relation to the complex voltage as \cite{milano2021complex}
\begin{equation}\label{compl_angl}
\underline{\vartheta}\coloneqq u+j\theta \quad \Rightarrow \quad \underline{v}=e^{\underline{\vartheta}}.
\end{equation}
The resulting relationship $\underline{v}=e^{\underline{\vartheta}}$ can be seen as a coordinate transformation between the complex-voltage coordinates $\underline{v}$ and the complex-angle coordinates $\underline{\vartheta}$. Considering the time-derivative of the complex angle in \eqref{compl_angl}, we can define the complex frequency as \cite{milano2021complex} 
\begin{equation}\label{compl_freq}
\underline{\varpi}\coloneqq\Dot{\underline{\vartheta}}=\frac{\dot{\underline v}}{\underline v}=\frac{\Dot{v}}{v}+j\Dot{\theta}=:\varepsilon+j\omega,
\end{equation}
where $\omega$ is the angular frequency and $\varepsilon$ the normalized rate of change of voltage (rocov). Therefore, the concept of complex frequency allows us to represent two-dimensional frequency information, i.e., the rocov in the radial direction and the angular frequency in the rotational direction, beyond the notion of the classical angular frequency with an underlying assumption of constant voltage magnitude. 

\subsection{Complex Power and Normalized Complex Power}
Complex power is a well-known compact representation of active power and reactive power, which is defined as
\begin{equation}
    s \coloneqq p + jq \coloneqq \underline v\, \underline {\overline i}_o,
\end{equation}
where $\underline v$ is the aforementioned complex voltage and $\underline {\overline i}_o$ the conjugate of complex current $\underline i_o$. Based on complex power, a normalized complex power variable has been defined and utilized in complex droop control to indicate its power balance. Specifically, the normalized complex power is defined as
\begin{equation}\label{norm_compl_pow}
\underline{\varsigma}\coloneqq\rho+j\sigma\coloneqq \frac{\underline{s}}{v^2}=\frac{p+jq}{v^2},
\end{equation}
where $\rho \coloneqq p/v^2$ is the normalized active power and $\sigma \coloneqq q/v^2$ the normalized reactive power. 
It has been shown in \cite{he2024complex} that in a complex-frequency synchronous state, the normalized complex power flow is invariant. We further defined a conjugated version of the normalized power as
\begin{equation}\label{eq:norm_compl_pow_conj}
    \underline{\overline \varsigma}\coloneqq\rho - j\sigma = \frac{p - jq}{v^2} = \frac{\underline{i}_o}{\underline v},
\end{equation}
which is utilized in the formulation of complex droop control.

\subsection{Static Complex Droop Control}
GFM control has been a preferred solution for the control of converter-dominated power systems. One of the existing GFM controls is the dispatchable virtual oscillator control (dVOC), which is an advanced version of virtual oscillator control (VOC) \cite{seo2019dispatchable}. 

The voltage behavior of a converter controlled by a dVOC is given in complex-voltage coordinates as
\begin{equation}\label{dVOC_1}
\underline{\dot{v}}=\underline{\varpi}_{0}\underline{v}+\eta e^{j\varphi}(\underline{\overline{\varsigma}}^{\star}\underline{v}-\underline{i}_{o})+\eta\alpha\frac{v^{\star}-v}{v^{\star}}\underline{v},
\end{equation}
where $\underline{\varpi}_{0} \coloneqq j \omega_0$ is the nominal complex frequency, $\underline{\overline{\varsigma}}^{\star}=(p^{\star}-jq^{\star})/{v^{\star}}^2$ the conjugated normalized power setpoint, $\underline{i}_{o}$ the converter output current, and $v^{\star}$ the voltage setpoint. Furthermore, the angle $\varphi\in[0,\pi/2]$ denotes the rotation of the current feedback to adapt to the resistive-inductive network impedance characteristics, and $\eta > 0$, $\alpha \geq 0$ are tunable control gains. Using \eqref{compl_freq} and \eqref{eq:norm_compl_pow_conj}, we can rewrite the dVOC in \eqref{dVOC_1} into a complex-power droop control in complex-frequency coordinates (i.e., complex droop control) as
\begin{equation}\label{dVOC_2}
\underline{\varpi}=\underline{\varpi}_0 + \eta e^{j\varphi}\left( \underline{\overline{\varsigma}}^{\star}-\underline{\overline{\varsigma}}+\alpha e^{-j\varphi}\frac{v^{\star}-v}{v^{\star}}\right).
\end{equation}
The mapping from the feedback $\underline{\overline{\varsigma}}$ and $v$ to the output $\underline{\varpi}$ is linear. By separating the complex frequency $\underline{\varpi}$ into real and imaginary parts, we can write the full expression of the static complex droop control in small-signal form as 
\begin{equation}\label{eq:static_complex_droop_ctrl}
\begin{split}
\hspace{-1mm}\begin{bmatrix}\Delta\varepsilon\\\Delta\omega\end{bmatrix}\hspace{-1mm}=\hspace{-1mm}\begin{bmatrix}\eta \cos{\varphi}&-\eta\sin{\varphi}\\\eta\sin{\varphi}&\eta \cos{\varphi}\end{bmatrix}\hspace{-1mm}
\left(\begin{bmatrix}-\Delta\rho\\\Delta\sigma\end{bmatrix}\hspace{-1mm} - \hspace{-1mm} \begin{bmatrix}\alpha\cos{\varphi}\\-\alpha\sin{\varphi}\end{bmatrix} \hspace{-1mm}\frac{\Delta v}{v^{\star}} \right)\hspace{-1mm},\hspace{-1mm}
\end{split}
\end{equation}
where $\Delta\varepsilon=\varepsilon-0$ and $\Delta\omega=\omega-\omega_0$ denote the rocov and the angular frequency deviations, from their respective nominal points. Furthermore, $\Delta\rho \coloneqq \rho - \rho^{\star}$, $\Delta\sigma \coloneqq \sigma - \sigma^{\star}$, and $\Delta v \coloneqq v - v^{\star}$ are normalized active power, reactive power, and voltage deviating from their respective setpoints.

\section{Dynamic Complex Frequency Control}\label{sec:dynamic_complex_freq_control}

\subsection{Single-Converter Control Setup}

\begin{figure}[t!]
     \centering
     \scalebox{1}{\includegraphics{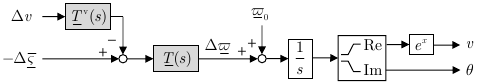}}
     \vspace{-2mm}
     \caption{Dynamic complex-frequency control architecture of a GFM converter.}
     \vspace{-3mm}
     \label{single_converter_control}
\end{figure}

Instead of using constant gain matrices as in the static complex droop control in \eqref{eq:static_complex_droop_ctrl}, the complex-frequency control can be formulated more generally in terms of \textit{dynamic} gain matrices (i.e., utilizing transfer functions) as
\begin{equation}\label{droop_dynamic}
\hspace{-1mm}\begin{bmatrix}\Delta\varepsilon\\\Delta\omega\end{bmatrix}\hspace{-1mm}=\hspace{-1mm}\begin{bmatrix}T^{\mathrm{re}}(s)&-T^{\mathrm{im}}(s)\\T^{\mathrm{im}}(s)&T^{\mathrm{re}}(s)\end{bmatrix} \hspace{-1mm}
\bigg(\hspace{-1mm}\begin{bmatrix}-\Delta\rho\\ \Delta\sigma\end{bmatrix}\hspace{-1mm}-\hspace{-1mm}\begin{bmatrix}T^{\rm vp}(s)\\T^{\rm vq}(s)\end{bmatrix} \hspace{-1mm}\Delta v \bigg),\hspace{-1mm}
\end{equation}
which can be more compactly written as
\begin{equation}\label{compact_droop}
\Delta\underline{\varpi}=\underline{T}(s)\Big(-\Delta\underline{\overline{\varsigma}}-\underline{T}^{\rm v}(s)\Delta v \Big),
\end{equation}
where $\underline{T}(s)\hspace{-0.5mm}\coloneqq\hspace{-0.5mm} T^{\mathrm{re}}(s) \hspace{-0.1mm}+\hspace{-0.1mm} j T^{\mathrm{im}}(s)$ and $\underline{T}^{\rm v}(s)\hspace{-0.5mm}\coloneqq \hspace{-0.5mm}T^{\rm vp}(s) \hspace{-0.1mm}+\hspace{-0.1mm} j T^{\rm vq}(s)$ are complex transfer functions that describe the corresponding control dynamics in GFM converters. The resulting dynamic complex-frequency control architecture is shown in Fig.~\ref{single_converter_control}, where, based on the definition of complex frequency, the voltage magnitude $v$ is given by $v \coloneqq e^{u}=e^{\int \varepsilon}$.

\subsection{Multi-Converter Control Setup}
We consider a more general system setup, where there are $n$ converters and they are collected together and connected to the main grid, as shown in Fig.~\ref{multi-converter-setup}. By applying Kron reduction, we can obtain a reduced model of the collector network, in which the intermediate nodes are eliminated \cite{haberle2023grid}. The admittance matrix of the reduced network is represented as $\underline{Y}_{\rm net}\in\mathbb{C}^{n \times n}$, where the admittance between the converter nodes $k$ and $l$ is described by the opposite of the entries $\underline{y}_{kl}$, respectively. The normalized power-flow equations and the linearly approximated ones (i.e., complex dc power flows) of the reduced network in Fig.~\ref{multi-converter-setup} are given as \cite{he2024complex}
\begin{equation}\label{norm_pow_conj}
    \underline{\overline{\varsigma}}_{{\rm e},k} = \sum\nolimits_{l=1}^{n}\underline{y}_{kl}\frac{\underline{v}_{l}}{\underline{v}_{k}}=\sum\nolimits_{l=1}^{n}\underline{y}_{kl}e^{\underline{\vartheta}_{l}-\underline{\vartheta}_{k}},
\end{equation}
\begin{equation}\label{norm_pow_dc}
    \underline{\overline{\varsigma}}_{\rm e}^{\rm dc} = \underline{Y}_{\mathrm{net}}\underline{\vartheta},
\end{equation}

\subsubsection{Aggregated Specification}
Each converter is controlled by a dynamic complex-frequency control as in \eqref{compact_droop}, where we now denote the local converter transfer functions with index $k$ as $\underline{T}_k(s)$ and $\underline{T}_k^\mathrm{v}(s)$, accordingly. By maintaining the same structure as in \eqref{compact_droop}, we specify a desired dynamic behavior of the multi-converter aggregation as 
\begin{equation}\label{compact_compl_freq_des_beh}
\Delta\underline{\varpi}_{\mathrm{pcc}}=\underline{T}_{\mathrm{des}}(s)\Big(\Delta\underline{\overline{\varsigma}}_{\mathrm{pcc}}-\underline{T}_{\mathrm{des}}^{\rm v}(s)\Delta v_{\mathrm{pcc}}\Big),
\end{equation}
where $\underline{T}_{\mathrm{des}} \coloneqq T_{\mathrm{des}}^{\mathrm{re}}(s) + j T_{\mathrm{des}}^{\mathrm{im}}(s)$ and $\underline{T}_{\mathrm{des}}^{\rm v}(s) \coloneqq T_{\mathrm{des}}^{\rm vp}(s) + j T_{\mathrm{des}}^{\rm vq}(s)$ describe the desired dynamics of the converter aggregation, which can be obtained from grid code specifications \cite{haberle2024dynamic}. Moreover, $\Delta v_{\mathrm{pcc}}$ is the measured PCC voltage deviation (deviating from its setpoint), $\Delta\overline{\underline{\varsigma}}_{\mathrm{pcc}} \coloneqq \Delta\rho_{\mathrm{pcc}} - j\Delta\sigma_{\mathrm{pcc}}$ with $\Delta\rho_{\mathrm{pcc}}$ and $\Delta\sigma_{\mathrm{pcc}}$ the PCC power disturbance (deviating from the setpoint, flowing into the converters), and $\Delta\varepsilon_{\mathrm{pcc}}$ and $\Delta\omega_{\mathrm{pcc}}$ the associated rocov and angular frequency deviations at the PCC (deviating from their respective nominal points).

\subsubsection{Aggregated Dynamics} 
\begin{figure}[t!]
     \centering
     \scalebox{1}{\includegraphics{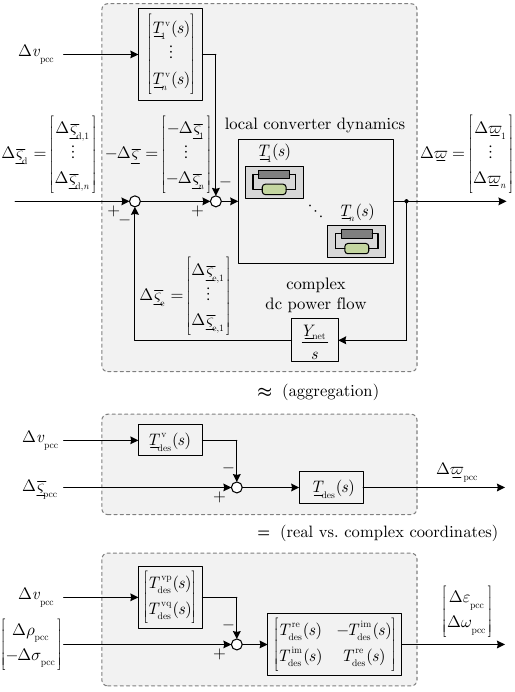}}
     \vspace{-2mm}
     \caption{Multi-converter control setup for dynamic complex-frequency control.}
     \vspace{-3mm}
     \label{multi_conv_dyn}
\end{figure}

The multi-converter dynamic complex-frequency control setup is illustrated in Fig.~\ref{multi_conv_dyn}. The input signals $\Delta\overline{\underline{\varsigma}}_{\mathrm{d},k}$ represent the local power disturbances at each converter node $k$, and the output signals $\Delta\underline{\varpi}_{k}$ the associated local complex-frequency deviations from its nominal value. The feedback dynamics described by ${\underline{Y}_{\mathrm{net}}}/{s}$ represent the complex dc power flow changes between the converters in the reduced network, cf. \eqref{norm_pow_dc}. More specifically, as further outlined in \cite{he2024complex}, the complex dc power flow is valid when the complex-angle differences between the converters are small (i.e., small voltage magnitude and phase differences), which is the case when providing dynamic ancillary services. 
\begin{figure}[t!]
     \centering
     \scalebox{0.95}{\includegraphics{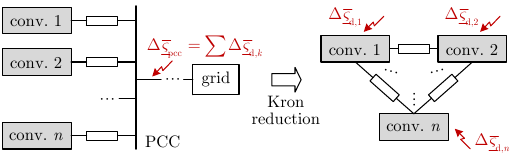}}
     \vspace{-2mm}
     \caption{A multi-converter network and its Kron reduction for control design.}
     \label{multi-converter-setup}
\end{figure}

From Fig.~\ref{multi_conv_dyn}, we describe the closed-loop dynamics as
\begin{equation}\label{dynamics2}
\hspace{-1mm}\Delta \overline{\underline{\varsigma}}_{\mathrm{d}}-\frac{\underline{Y}_{\mathrm{net}}}{s}\Delta\underline{\varpi}-\hspace{-1mm}\begin{bmatrix}\underline{T}_{1}^{\rm v}(s)\\\vdots\\\underline{T}_{n}^{\rm v}(s)\end{bmatrix}\hspace{-1mm}\Delta v_{\mathrm{pcc}}=\mathrm{diag}\left({\underline{T}_{k}^{-1}(s)}\right)\hspace{-0.5mm}\Delta\underline{\varpi}. 
\end{equation}
Next, we consider that all the converters already exhibit a synchronized response (the synchronization/stability guarantee will be shown later), which is denoted as $\Delta\underline{\varpi} = 1_n \Delta\underline{\varpi}_{\mathrm{sync}}$. Based on this, by additionally left-multiplying both sides of \eqref{dynamics2} with the vector of all ones $1_n^{\mathsf{T}}$, we obtain
\begin{multline*}
    1_n^{\mathsf{T}}\Delta \overline{\underline{\varsigma}}_{\mathrm{d}}-\tfrac{1_n^{\mathsf{T}}\underline{Y}_{\mathrm{net}}}{s} 1_n \Delta\underline{\varpi}_{\mathrm{sync}}-\textstyle \sum_{k=1}^{n}\underline{T}_k^{\rm v}(s) \Delta v_{\mathrm{pcc}}=\\
    \textstyle\sum_{k=1}^{n}\underline{T}_k^{-1}(s)\Delta\underline{\varpi}_{\mathrm{sync}},
\end{multline*}
where $1_n^{\mathsf{T}}\underline{Y}_{\mathrm{net}}=0$, by definition of the admittance matrix. Thus, assuming synchronization, the synchronous complex frequency at the PCC can be written as
\begin{equation}\label{eq:dynamics4}
\hspace{-1mm}\Delta\underline{\varpi}_{\mathrm{pcc}}\hspace{-0.5mm}=\hspace{-0.5mm}\Delta\underline{\varpi}_{\mathrm{sync}}\hspace{-0.5mm}=\hspace{-0.5mm}\frac{\sum\nolimits_{k=1}^{n}\Delta \overline{\underline{\varsigma}}_{\mathrm{d},k}-\sum\nolimits_{k=1}^{n}\underline{T}_k^{\rm v}(s) \Delta v_{\mathrm{pcc}}}{\sum\nolimits_{k=1}^{n}\underline{T}_k^{-1}(s)}.\hspace{-1mm}
\end{equation}
Utilizing the lossless property of the dc power flow \cite{he2024complex}, we obtain $\sum\nolimits_{k=1}^n\Delta \overline{\underline{\varsigma}}_{\mathrm{d},k}\approx\Delta\overline{\underline{\varsigma}}_{\mathrm{pcc}}$ as shown in Fig.~\ref{multi-converter-setup}. Therefore, \eqref{eq:dynamics4} further simplifies to
\begin{equation}\label{dynamics5}
\Delta\underline{\varpi}_{\mathrm{pcc}}=\frac{\Delta\overline{\underline{\varsigma}}_{\mathrm{pcc}}-\sum\nolimits_{k=1}^{n}\underline{T}_k^{\rm v}(s) \Delta v_{\mathrm{pcc}}}{\sum\nolimits_{k=1}^{n}\underline{T}_k^{-1}(s)}.
\end{equation}
By matching the expression \eqref{dynamics5} with the desired aggregate behavior in \eqref{compact_compl_freq_des_beh}, the aggregation conditions for complex-frequency control are finally obtained as
\begin{equation}\label{agg_cond_compl}
    \hspace{-2mm}\Bigl(\sum\nolimits_{k=1}^{n} \underline{T}_k^{-1}(s)\Bigr)^{-1}\hspace{-0.5mm}\stackrel{!}{=}\hspace{-0.5mm}\underline{T}_{\mathrm{des}}(s),\ \sum\nolimits_{k=1}^{n}\underline{T}^{\rm v}_k(s)\hspace{-0.5mm}\stackrel{!}{=}\hspace{-0.5mm}\underline{T}^{\rm v}_{\mathrm{des}}(s),\hspace{-1mm}
\end{equation}
where $``\stackrel{!}{=}"$ indicates desired equality to be satisfied. Given the desired aggregate behavior of $\underline{T}_{\mathrm{des}}(s)$ and $\underline{T}^{\rm v}_{\mathrm{des}}(s)$, the control design problem is to find local converter controllers $\underline{T}_k(s)$ and $\underline{T}^{\rm v}_k(s)$, such that the conditions \eqref{agg_cond_compl} are satisfied.

\subsubsection{Disaggregation of the Desired Behavior}
To obtain the local controller transfer functions of each converter, we can follow a divide-and-conquer strategy, similarly as introduced in \cite{haberle2021control}. More specifically, we disaggregate the transfer functions $\underline{T}_{\mathrm{des}}(s)$ and $\underline{T}^{\rm v}_{\mathrm{des}}(s)$ as
\begin{equation}\label{disaggregation}
\begin{split}
&\underline{T}_{\mathrm{des}}^{-1}(s)=\sum\nolimits_{k=1}^{n}m_k(s)\underline{T}_{\mathrm{des}}^{-1}(s)\stackrel{!}{=}\sum\nolimits_{k=1}^{n}\underline{T}_k^{-1}(s), \\
&\underline{T}^{\rm v}_{\mathrm{des}}(s)=\sum\nolimits_{k=1}^{n}m_k^{\rm v}(s)\underline{T}^{\rm v}_{\mathrm{des}}(s)\stackrel{!}{=} \sum\nolimits_{k=1}^{n}\underline{T}^{\rm v}_k(s),
\end{split}
\end{equation}
where the transfer functions $m_k(s)$ and $m_k^{\rm v}(s)$ are so-called dynamic participation factors, which have to satisfy
\begin{equation}\label{part_conditions}
\sum\nolimits_{k=1}^{n}m_k(s)\stackrel{!}{=}1, \quad \sum\nolimits_{k=1}^{n}m_k^{\rm v}(s)\stackrel{!}{=}1.
\end{equation}
The participation factors are selected such that the conditions in \eqref{part_conditions} are satisfied, while simultaneously accounting for the timescale and capacity limitations of the individual converter sources (see \cite{haberle2021control,haberle2023grid} for further details). Based on the disaggregation in \eqref{disaggregation}, we finally obtain the transfer functions of the local converter controller as 
\begin{equation}\label{loc_cond_compl}
\hspace{-1mm}
\underline{T}_k(s){=}m_k^{-1}(s)\underline{T}_{\mathrm{des}}(s),\  \underline{T}_k^{\rm v}(s){=}m_k^{\rm v}(s)\;\underline{T}_{\mathrm{des}}^{\rm v}(s),\,\forall k
\end{equation}
The implementation of the local controller transfer functions adopts the same control structure as in Fig.~\ref{single_converter_control}.

The multi-converter control setup has been further extended to allow for fully flexible desired responses \cite{domingo2023complex}, with $\underline{T}_{\mathrm{des}}$ generalized to arbitrary $2 \times 2$ transfer matrices. Moreover, the control design is valid for general network characteristics with nonuniform $X/R$ ratios. These benefits cannot be achieved using separate frequency and voltage control setups \cite{haberle2023grid}.

\subsection{Small-Signal Stability Guarantees} The use of complex-frequency coordinates in the control design allows us to obtain stability guarantees for the closed-loop system in Fig.~\ref{multi_conv_dyn} in terms of both frequency synchronization and voltage stabilization. Compared to the prior result in \cite{jiang2021grid,paganini2019global}, where the stability of the system was obtained under the assumption of fixed voltages, our stability result in this work inherently considers both frequency and voltage dynamics.

For ease of analysis, we assume that the collector network has a uniform $R/X$ ratio, to which the corresponding consistent impedance angle is denoted by $\varphi_z$. This allows us to simplify the dc power flow feedback as $\frac{1}{s}\underline{Y}_{\rm net} = \frac{1}{s}e^{-j\varphi_z}L_{\rm net}$, where $L_{\rm net}$ is real-valued and Laplacian, representing the element-wise magnitude of $\underline{Y}_{\rm net}$. Thus, the closed-loop system in Fig.~\ref{multi_conv_dyn} is equivalent to a closed-loop connection of $\mathrm{diag}\left({e^{-j\varphi_z} \underline{T}_{k}(s)}\right)$ and $\frac{1}{s}L_{\rm net}$, where $\Delta \overline{\underline{\varsigma}}_{\mathrm{d}}$ and $\Delta v_{\mathrm{pcc}}$ are considered as exogenous disturbances. Since $\frac{1}{s}L_{\rm net}$ is always positive real (i.e., passive), it is straightforward to find that if the equivalent controller $e^{-j\varphi_z} \underline{T}_{k}(s)$ is strictly positive real (i.e., strictly passive), then the closed-loop system is internally stable \cite[Proposition 1]{paganini2019global}. Moreover, if the voltage controller $\underline{T}_{k}^{\rm v}(s)$ is asymptotically stable, then the feedforward path from $\Delta v_{\mathrm{pcc}}$ to $\Delta \underline{\varpi}$ is also internally stable.

If using the local voltage $\Delta v_{k}$ to replace the PCC voltage $\Delta v_{\rm pcc}$ in the control design, we can also obtain such stability guarantees. Specifically, the local voltage feedback can be rewritten as $\Delta v_{k} = \frac{v_{k,0}}{s} \Re\{\Delta \underline{\varpi}_k\}$, and thus $\underline{T}_{k}^{\rm v}(s)\frac{v_{k,0}}{s} $ can be seen as another feedback connection from the output $\Re\{\Delta \underline{\varpi}_k\}$ to the input $\Delta \overline{\underline{\varsigma}}_{\mathrm{d}}$, in parallel to $\frac{1}{s}\underline{Y}_{\rm net}$, with the equivalent entire feedback as $\mathrm{diag}\left(e^{j\varphi_z}\underline{T}_{k}^{\rm v}(s) \frac{v_{k,0}}{s}\right) + \frac{1}{s}L_{\rm net}$. Therefore, if the dynamics $e^{j\varphi_z}\underline{T}_{k}^{\rm v}(s) \frac{v_{k,0}}{s}$, in addition to $\frac{1}{s}L_{\rm net}$, are passive, the closed-loop stability will also be guaranteed.

\section{Case Studies}\label{sec:case_studies}
\begin{figure}
     \centering
          \vspace{-3mm}
     \scalebox{0.5}{\includegraphics{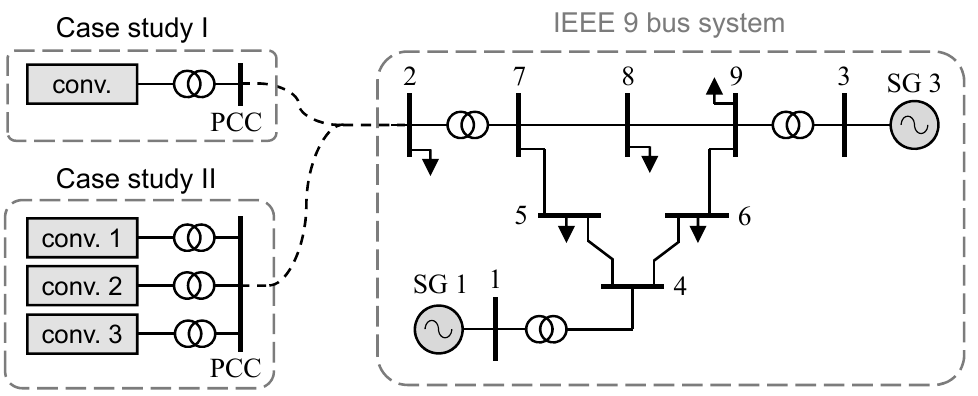}}
     \vspace{-2mm}
     \caption{IEEE nine-bus system with dynamic complex-frequency controlled converters at bus 2.}
     \label{fig:case_study_system}
\end{figure}

To verify the dynamic complex-frequency control strategy, we use Simscape Electrical in Matlab/Simulink to perform electromagnetic transient simulations on the IEEE nine-bus system (Fig.~\ref{fig:case_study_system}). More specifically, in a first case study, we show the performance of the dynamic complex-frequency control implemented in a single GFM converter system that replaces the classical frequency and voltage control of a thermal-based generator. For this setup, we also demonstrate the superiority of the dynamic complex-frequency control over the static complex droop control. Afterwards, in a second case study, we investigate the performance of the dynamic complex-frequency control when applied in a multi-converter setup.

\subsection{System Model}
The implementation of the IEEE nine-bus system in Fig. \ref{fig:case_study_system} is mainly based on the system and device models presented in \cite{haberle2021control}, where the transmission lines are modeled via nominal $\pi$ sections, and the transformers via three-phase linear models. As in \cite{haberle2021control}, for the synchronous generators (SGs) at buses 1 and 3, we adopted an 8th-order model equipped with an ST1A excitation system that includes a built-in automatic voltage regulator and a power system stabilizer. The governors are modeled as proportional speed-droop control with first-order delay, and the steam turbine parameters are from \cite{kundur2007power}.

\subsection{Grid-Forming Converter Model}\label{sec:converter}
\begin{figure}[t!]
     \centering
\scalebox{0.42}{\includegraphics{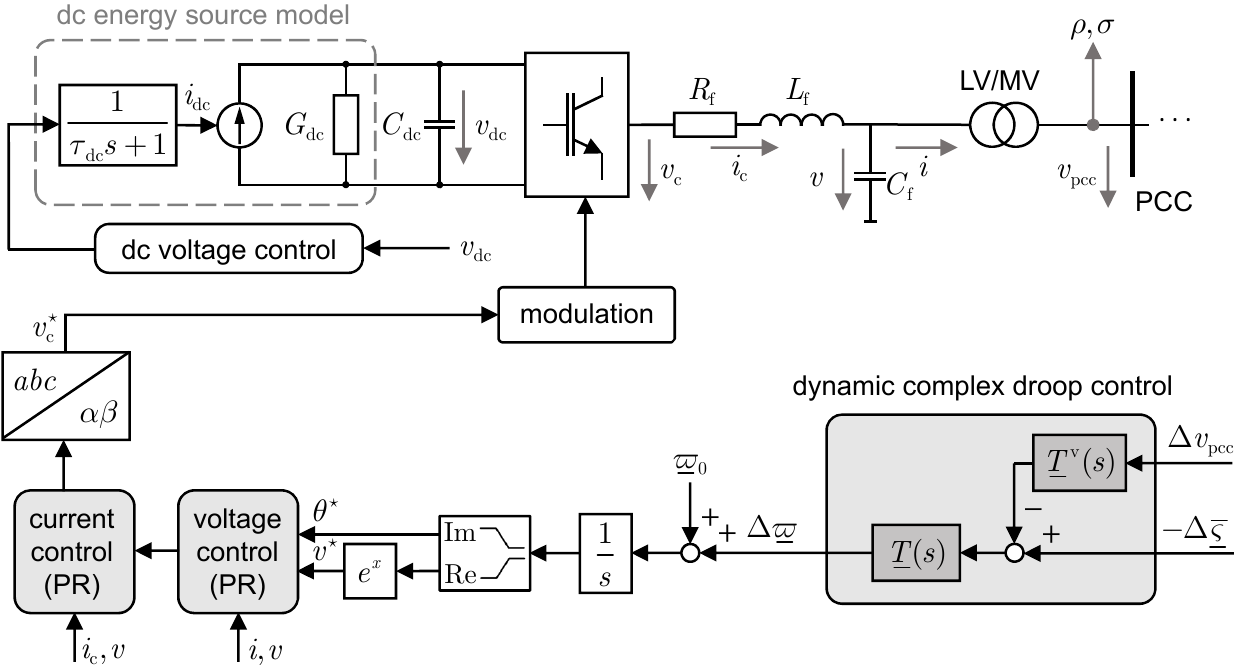}}
     \vspace{-3mm}
     \caption{Block diagram of the dynamic complex-frequency control and inner control loops in a single-converter setup.}
     \vspace{-3mm}
     \label{gfm_converter_model}
\end{figure}

For both case studies, we adopt a uniform GFM converter topology as illustrated in Fig.~\ref{gfm_converter_model}. The relevant converter parameters are listed in Table~\ref{tab:converter}. The converter system represents an aggregation of multiple commercial converter modules. Following a similar approach to \cite{tayyebi2020frequency}, the primary energy source on the converter's DC side is modeled as a generic controllable DC current source, exhibiting a delayed response characteristic to account for primary source dynamics and potential actuation delays. The converter system connects to the AC grid via an $RLC$ filter and an LV/MV transformer with a rating of 1/18\,kV (phase-phase, RMS). The AC side control of the converter is structured according to a standard hierarchical control architecture, with the proposed dynamic complex-frequency control strategy incorporated into the outer loop to generate reference signals for the cascaded inner-loop controllers (implemented in $\alpha\beta0$ coordinates).

\begin{table}[t!]
\caption{Converter parameters}\label{tab:converter}
\vspace{-3mm}
\centering
\setlength\tabcolsep{2.5pt} 
\begin{tabular}{c||c|c}
 \hline
\multicolumn{3}{c}{Converter parameters rated at $S_{r}$, $v_r=\sqrt{2/3}$\,kV (ph-n peak), $v_{\mathrm{dc}}^{\star}=3v_r$
} \\
 \hline
 parameter& symbol & value\\
 \hline
 dc-link components& $C_{\mathrm{dc}}$, $G_{\mathrm{dc}} $ &0.096\,pu, 0.05\,pu \\
 $RLC$ filter components&$R_f$,$L_f$,$C_f$& 0.01\,pu, 0.11\,pu, 0.0942\,pu\\
 LV/MV transformer components&$R_t$, $L_t$& 0.01\,pu, 0.1\,pu\\
 dc-source time constant&$\tau_{\mathrm{dc}}$&0.01\,s\\
 \hline
\end{tabular}
\end{table}

\subsection{Case Study I: Comparison of Dynamic Complex-Frequency Control With Static Complex Droop Control}
For the GFM converter model introduced above, we specify the dynamic complex-frequency control behavior in \eqref{compact_droop} as
\begin{equation}\label{case1_dyn}
\underline{T}(s)=\tfrac{e^{j\varphi}}{M_ds+D_d}, \quad \underline{T}^{\rm v}(s)= \alpha_d e^{-j\varphi},
\end{equation}
where $D_d=50$ and $\alpha_d=5$ are droop gains, $M_d=2$\,s an inertia time constant, and $\varphi = \pi/4$ to adapt to the entire network impedance characteristics. These parameters are chosen to provide inertial response and steady-state drooped operation. In the following, we will demonstrate the superiority of the dynamic complex-frequency control over the static complex droop control. To get a fair comparison we select equivalent droop gains for both control strategies, i.e., we select the coefficients of the static complex droop control in \eqref{dVOC_2} as $\eta_s=1/D_d=0.02$ and $\alpha_{s}=5$, accordingly.

\begin{figure}[t!]
    \centering
    \scalebox{0.58}{\includegraphics{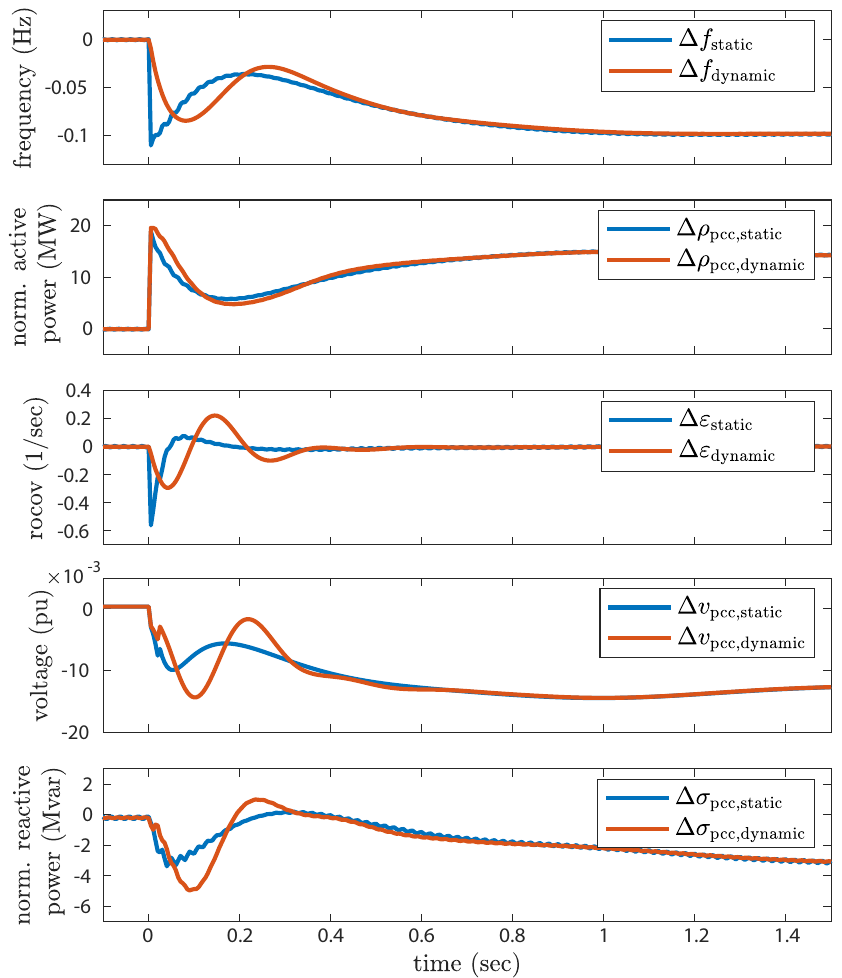}}
    \vspace{-2mm}
    \caption{System response of the dynamic complex-frequency control and the static complex droop control in case study I during a
 load increase at bus 2.}
 \vspace{-3mm}
    \label{fig:case1_plots}
\end{figure}

We simulate a load increase of 25 MW at bus 2 and investigate the system response behavior of both static complex droop and dynamic complex-frequency control at the PCC. Specifically, as depicted by the red curves in Fig.~\ref{fig:case1_plots}, dynamic complex-frequency control exhibits an inertial dynamic response in both frequency and rocov, aligning with the specifications in \eqref{case1_dyn}. In contrast, static complex droop control demonstrates significantly poorer behavior, resulting in a sudden change in complex frequency signals (blue lines), attributed to the absence of inertial response specification. We note that the voltage response exhibit an oscillatory response, which is because the inertial $\underline{T}(s)$ has also been applied to the rocov. To allow for non-inertial rocov compared to inertial frequency behaviors, one can specify fully flexible desired responses as described in \cite{domingo2023complex}.

\subsection{Case Study II: Dynamic Complex-Frequency Control in a Multi-Converter Setup}
In this case study, we examine an aggregation of three GFM converters at bus 2, depicted in Fig.~\ref{fig:case_study_system}. We focus on a homogeneous group of converter systems, following the model outlined in Section~\ref{sec:converter}, characterized by identical primary source characteristics. Specifically, we intentionally employ such a small and homogeneous group of converters to illustrate the fundamental concept of our multi-converter control strategy. Nevertheless, our approach can be readily expanded to accommodate more intricate setups, encompassing a greater number of converter units and/or diverse primary source technologies with different response characteristics.

Similar to case study I, we specify the desired aggregate behavior at the PCC in \eqref{compact_compl_freq_des_beh} as
\begin{equation}
\underline{T}_{\mathrm{des}}(s)= \tfrac{e^{j\varphi}}{M_ds+D_d}, \quad \underline{T}_{\mathrm{des}}^{\rm v}(s)= \alpha_d e^{-j\varphi},
\end{equation}
where $M_d=2$\,s, $D_d=50$, $\alpha_{d}=5$, and $\varphi = \pi/4$. Given the homogeneous group of converters, we choose equal participation factors as $m_k(s)=1/3$ and $m_k^{\rm v}(s)=1/3$. 

\begin{figure}[t!]
    \centering
    \scalebox{0.58}{\includegraphics{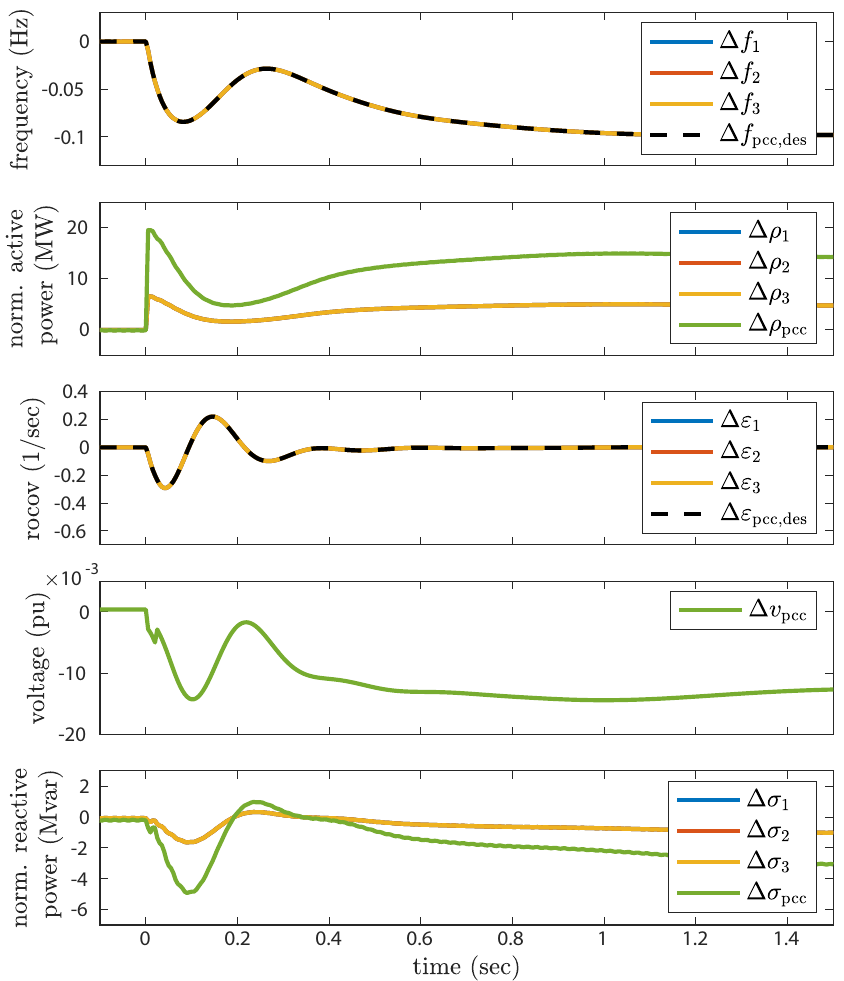}}
    \vspace{-2mm}
    \caption{System response of the complex frequency controlled converter aggregation in case study II during a load increase at bus 2.}
    \vspace{-3mm}
    \label{fig:case2_plots}
\end{figure}

We simulate a 25 MW load increase on bus 2 and investigate the complex frequency response of the converter aggregation at the PCC. Fig.~\ref{fig:case2_plots} shows how all the converters have a synchronized complex frequency response. This satisfies the assumption of a synchronized complex frequency response of all the converters. In addition, the converter responses precisely match the specified desired complex frequency at the PCC (dashed lines). Moreover, given identical participation factors, all the converters come with the same power injections. 

\section{Conclusion}\label{sec:conclusion}
We have proposed the novel concept of dynamic complex-frequency control, a GFM converter control strategy that extends the recently developed (static) complex droop control by incorporating dynamism. Specifically, this involves upgrading the static droop gains with dynamic transfer functions, thereby enhancing richness and flexibility to swiftly respond to dynamic changes in frequency and voltage, crucial for modern power systems and next-generation grid codes. Unlike existing methods, our approach effectively manages the coupling between frequency and voltage dynamics, ensuring stability guarantees for frequency synchronization and voltage stabilization. The dynamic complex-frequency control framework is also able to address fully flexible dynamic response specifications. 

Future research will explore heterogeneous primary source technologies within multi-converter setups, as well as dynamic interaction stability with the power grid.

\bibliographystyle{IEEEtran}
\bibliography{IEEEabrv,bibliography}

\end{document}